\documentclass[aps,prb,reprint,english,a4paper,10pt,floatfix,superscriptaddress,showpacs]{revtex4-1}
\usepackage[T1]{fontenc}
\usepackage{ae,aecompl}
\usepackage[english]{babel}
\usepackage[normalem]{ulem}
\usepackage{url}
\usepackage{psfrag,color,pstricks,pst-grad}
\usepackage[dvips]{graphicx}
\usepackage{amsmath,amsfonts,amssymb,amsbsy}  

\newcommand{\bb}{\mathbf}

\newcommand{\br}{\mathbf{r}}
\newcommand{\bv}{\mathbf{v}}

\newcommand{\begeq}{\begin{equation}}
\newcommand{\itoj}{j \to i} 

\begin{document}
 
  \title{Role of anharmonic phonon scattering in the spectrally decomposed \\ thermal conductance at planar interfaces}
\author{K. S\"a\"askilahti}
\email{kimmo.saaskilahti@aalto.fi}
\affiliation{Department of Biomedical Engineering and Computational Science, Aalto University, FI-00076 Aalto, Finland}
\affiliation{Ecole Centrale Paris, Grande Voie des Vignes, 92295 Ch\^atenay-Malabry, France}
\author{J. Oksanen}
\affiliation{Department of Biomedical Engineering and Computational Science, Aalto University, FI-00076 Aalto, Finland}
\author{J. Tulkki}
\affiliation{Department of Biomedical Engineering and Computational Science, Aalto University, FI-00076 Aalto, Finland}
\author{S. Volz}
\email{sebastian.volz@ecp.fr}
\affiliation{Ecole Centrale Paris, Grande Voie des Vignes, 92295 Ch\^atenay-Malabry, France}
\affiliation{CNRS, UPR 288 Laboratoire d'Energ\'etique Mol\'eculaire et Macroscopique, Combustion (EM2C), Grande Voie des Vignes, 92295 Ch\^atenay-Malabry, France}

\date{\today}
\pacs{05.60.Cd, 44.10.+i, 63.22.-m} 

\begin{abstract}
Detailed understanding of vibrational heat transfer mechanisms between solids is essential for the efficient thermal engineering and control of nanomaterials. We investigate the frequency dependence of anharmonic scattering and interfacial thermal conduction between two acoustically mismatched solids in planar contact by calculating the spectral decomposition of the heat current flowing through an interface between two materials. The calculations are based on analyzing the correlations of atomic vibrations using the data extracted from non-equilibrium molecular dynamics simulations. Inelastic effects arising from anharmonic interactions are shown to significantly facilitate heat transfer between two mass-mismatched face-centered cubic lattices even at frequencies exceeding the cut-off frequency of the heavier material due to (i) enhanced dissipation of evanescent vibrational modes and (ii) frequency-doubling and frequency-halving three-phonon energy transfer processes at the interface. The results provide substantial insight into interfacial energy transfer mechanisms especially at high temperatures, where inelastic effects become important and other computational methods are ineffective.




 \end{abstract}
 \maketitle
 
\section{Introduction}
 
Proper thermal management is one of the key factors determining the performance of future nanodevices \cite{pop10}. Because of the relatively long phonon mean free path in nanoscale structures,  phonons carrying the heat are typically primarily scattered at material interfaces and boundaries \cite{cahill03,cahill14}. The properties of interfacial transport therefore often play the key role in the determination of the thermal conductance of the composite structure. 
 
The fact that phonon scattering at the interface between two pure but dissimilar materials leads to thermal resistance was first observed by Kapitza \cite{kapitza41} for the interface between solid and liquid helium. The effect was explained by Khalatnikov \cite{khalatnikov52} and Little \cite{little59} in terms of the mismatch between the acoustic properties of the materials. This model is now known as the acoustic mismatch model (AMM) and, along with the diffuse mismatch model \cite{swartz89}, it has proven to be a useful rule-of-thumb in calculating the thermal interfacial resistance based on simple material properties.   

Because of its phenomenological nature, AMM cannot provide as detailed picture of interfacial thermal conduction as atomistic scattering methods such as the Green's function (GF) method \cite{mingo03,zhang07}. These methods rely, however, on the linear approximation of interatomic forces and cannot therefore directly describe inelastic effects. The limitation to harmonic transport is a serious drawback at high temperatures and for weakly-bonded interfaces, where inelastic effects become important \cite{shen11,chalopin12b}. In contrast to the GF method, molecular dynamics (MD) simulations can straightforwardly describe the inelastic effects through the anharmonic interatomic forces used in integrating the classical equations of motion \cite{allentildesley}. MD is, however, directly suitable only for calculating the total interfacial thermal resistance \cite{puech86,maiti97,mullerplathe97,jund99,barrat03,rajabpour10}, which cannot give a detailed, spectrally resolved picture of energy transfer processes at the interface. 

In this article, we investigate the frequency dependent contribution of inelastic scattering to interfacial thermal conduction by developing a new method to calculate the spectral decomposition of interfacial thermal conductance, offering a detailed description of inelastic effects. The atomic cross-correlation functions required to calculate the spectral decomposition are obtained from microscopic dynamics using non-equilibrium MD simulations. In an earlier work \cite{chalopin12,chalopin13} co-authored by one of the present authors, the interfacial equilibrium fluctuations were used to estimate the energy transmission function in investigating the contribution of resonant interfacial modes to the thermal conductance. In contrast to this earlier work, the developed method allows us to evaluate the actual spectral heat current at the interface from \textit{non-equilibrium} steady-state simulations subject to finite temperature differences.

We present numerical results for two non-linear, mass-mismatched face-centered cubic lattices set in perfect contact. Our results show that (i) anharmonic effects in the bulk and at the interface enable energy transmission at frequencies exceeding the cut-off frequency of the heavier material, (ii) frequency-doubling and frequency-halving processes dominate the inelastic energy transfer at the interface, and (iii) uniaxial stress enhances the elastic transmission of mid-wavelength phonons across the interface, while the inelastic transfer is largely unaffected by the stress. 


The paper is organized as follows. In Sec. \ref{sec:distribution}, we derive a general microscopic expression for the spectral heat current distribution, which forms the basis for the spectral decomposition of interfacial conductance. To identify the elastic and inelastic contributions, the heat current distribution is divided into the harmonic and anharmonic parts in Sec. \ref{sec:2B}, giving a final expression for the spectral conductance. The decomposition formula is applied to calculate the frequency-dependent interfacial conductance between two mass-mismatched Lennard-Jones solids using non-equilibrium molecular dynamics in Sec. \ref{sec:results}, where the effects of temperature and pressure on the inelastic processes are investigated. We conclude in Sec. \ref{sec:conclusions}. 
 
 \section{Theory}
 \label{sec:theory}
 
 \subsection{Spectral heat current distribution}
 \label{sec:distribution}
Thermal conductance $G$ across an interface is defined as 
 \begin{equation}
 G = \frac{Q}{A\Delta T} , \label{eq:Gdef}
 \end{equation}
where Q is the time-averaged, steady-state heat current across the interface, $A$ is the interfacial area and $\Delta T$ is the temperature change at the interface. Definition implicitly assumes the limit $\Delta T\to 0$ so that the heat current is linear in $\Delta T$. To access the spectrally resolved conductance, it is necessary to determine the spectrally resolved thermal current $q(\omega)$ defined through the relation
\begin{equation}
 Q = \int_0^{\infty} \frac{d\omega}{2\pi} q(\omega), \label{eq:q_def}
\end{equation}
where $\omega$ is the angular frequency and $q(\omega)$ is the \textit{deterministic, time-averaged} spectral heat current. Note that $q(\omega)$ is not directly related to the microscopic thermal fluctuations of the heat current itself. 



To derive an expression for the spectral distribution $q(\omega)$ of heat current between two atom sets $I$ and $J$ in thermal contact in non-equilibrium steady-state, we start from the general expression for the conduction current $Q_{\itoj}$ between any two atoms $i\in I$ and $j\in J$, given by \cite{hardy63,lepri03,saaskilahti12}
\begin{alignat}{2}
 Q_{\itoj}  &=  \frac{1}{2} \langle \bb{F}_{ij} \cdot (\bv_i +\bv_j) \rangle.
 \label{eq:Qij_def}
\end{alignat}
The average on the right-hand side of \eqref{eq:Qij_def} refers to the non-equilibrium ensemble average assumed to be equal to the time average due to ergodicity. Equation \eqref{eq:Qij_def} is essentially the work done on atom $i$ at position $\bb{r}_i$ by the interatomic force $\bb{F}_{ij}=-\bb{F}_{ji}$ acting on atom $i$ due to the atom $j$ at position $\bb{r}_j$. The force can be derived from the interatomic potential $V(\br_i-\br_j)$ as $\bb{F}_{ij}=-\partial V(\bb{r}_i-\br_j)/\partial \br_i$ and the velocity of atom $i$ is denoted by $\bv_i$. To calculate $q(\omega)$ of \eqref{eq:q_def}, we need to write also Eq. \eqref{eq:Qij_def} in the form
\begin{equation}
 Q_{\itoj} = \int_0^{\infty} \frac{d\omega}{2\pi} q_{\itoj}(\omega), \label{eq:qij_def}
\end{equation}
where $q_{\itoj}(\omega)$ is the interparticle spectral heat current. The spectral decomposition \eqref{eq:q_def} of the interfacial thermal current $Q$ can then be obtained by summing Eq. \eqref{eq:qij_def} over all atom pairs interacting across the interface.

The spectral decomposition of thermal current \eqref{eq:Qij_def} can be related to the correlation time between the force and velocity terms \cite{chalopin13}. We therefore define the auxiliary correlation function
\begin{equation}
 K_{ij}(t_1-t_2) =  \frac{1}{2}  \left\langle \bb{F}_{ij}(t_1)\cdot [\bv_i(t_2) +\bv_j(t_2)]  \right\rangle ,
 \label{eq:Jij1}
\end{equation} 
which depends only on the time difference $t_1-t_2$ due to the assumed steady state and ensemble averaging. The Fourier transform $\tilde K_{ij}(\omega)$ and the inverse transform are defined, as usual, as $\tilde K_{ij}(\omega) = \int_{-\infty}^{\infty} d \tau  e^{i\omega \tau} K_{ij}(\tau)$ and $K_{ij}(\tau) = \int_{-\infty}^{\infty} (d\omega/2\pi)  e^{-i\omega \tau} \tilde K_{ij}(\omega)$. Using the definition of the inverse transform and noting that $K_{ij}(0)\equiv Q_{\itoj}$, we see that
\begin{equation}
  Q_{\itoj}  = \int_{-\infty}^{\infty} \frac{d\omega}{2\pi} \tilde K_{ij}(\omega). \label{eq:Q_Kij}
\end{equation}
Since $K_{ij}(\tau)$ is real, the real and imaginary parts of $\tilde K_{ij}(\omega)$ are even and odd functions, respectively, and Eq. \eqref{eq:Q_Kij} further simplifies to
\begin{equation}
  Q_{\itoj}  =2 \int_{0}^{\infty} \frac{d\omega}{2\pi} \textrm{Re}[\tilde K_{ij}(\omega)].
 \label{eq:QAB_JAB}
\end{equation}
This form shows that the spectral heat current $q_{\itoj}(\omega)$ defined in Eq. \eqref{eq:qij_def} is 
\begin{equation}
 q_{\itoj}(\omega) = 2 \textrm{Re}[\tilde K_{ij}(\omega)].
 \label{eq:spectral_J}
\end{equation}
This equation for the spectral heat current defined in terms of the Fourier transform of \eqref{eq:Jij1} is our first main result. 

 \subsection{Elastic and inelastic energy transmission}
\label{sec:2B}
Equation \eqref{eq:spectral_J} can be used to determine the spectral distribution of interatomic heat conduction current in solid, liquid and gas systems from statistical data obtained, e.g., from non-equilibrium molecular dynamics. This requires, however, storing the force and velocity trajectories of all atom pairs participating in heat transfer on disk for Fourier transform analysis, as usual for correlation functions \cite{allentildesley}. In solids, the atoms remain close to their average positions and one can reduce the computational burden by expanding the interatomic forces in terms of small displacements from the average atomic positions. The expansion additionally provides separate expressions for elastic and many-phonon thermal conduction processes, offering more insight into the interfacial thermal conduction.

When the atoms are vibrating close to their average positions $\bb{r}_i^0=\langle \bb{r}_i \rangle$, one can expand the interatomic force $\mathbf{F}_{ij}$ in a Taylor series in terms of the small particle displacements $\bb{u}_i=\bb{r}_i-\bb{r}_i^0$ as
\begin{alignat}{2}
 F^{\alpha}_{ij} \approx  &\sum_{\beta \in\{x,y,z\}}  k_{ij} ^{\alpha\beta}\left(u_j^{\beta}-u_i^{\beta} \right) \notag \\
 &+\frac{1}{2} \sum_{\beta,\gamma \in\{x,y,z\}}  \gamma_{ij}^{\alpha \beta \gamma} \left(u_j^{\beta}-u_i^{\beta}\right) \left(u_j^{\gamma}-u_i^{\gamma}\right) , \label{eq:f_expansion}
\end{alignat}
where
\begin{equation}
k_{ij}^{\alpha \beta} = \left. \frac{\partial F_i^{\alpha}}{\partial u_j^{\beta} }\right|_{\bb{u}=0}
\end{equation}
and 
\begin{equation}
\gamma_{ij}^{\alpha\beta\gamma} = \left. \frac{\partial^2 F_i^{\alpha}}{\partial u_j^{\beta} \partial u_j^{\gamma}}\right|_{\bb{u}=0}
\end{equation}
are, respectively, the harmonic and first-order anharmonic interatomic force constants. Substituting Eq. \eqref{eq:f_expansion} to Eq. \eqref{eq:Jij1}, we get 
\begin{alignat}{2}
 K_{ij} (\tau) \approx & \frac{1}{2}\sum_{\alpha,\beta \in\{x,y,z\}}  k_{ij} ^{\alpha\beta} A^{\alpha\beta}_{ij}(\tau) \notag \\
 & + \frac{1}{4}\sum_{\alpha,\beta,\gamma \in\{x,y,z\}} \gamma_{ij}^{\alpha\beta\gamma} B_{ij}^{\beta\gamma\alpha}(0,\tau)   , \label{eq:Q_ABtau}
\end{alignat}
where the approximation sign stems from the truncation of the force expansion \eqref{eq:f_expansion} after the second term and the correlation functions $A_{ij}^{\beta\alpha}(t)$ and $B_{ij}^{\beta\gamma\alpha}(t,t')$ are defined as
\begin{equation}
A_{ij}^{\beta\alpha}(t_1-t_2) = \left\langle \left[u_j^{\beta}(t_1)-u_i^{\beta}(t_1) \right]\left[\vphantom{u_j^{\beta}(t_1)} v_i^{\alpha}(t_2)+v_j^{\alpha}(t_2) \right] \right\rangle \label{eq:Adef} 
\end{equation}
and 
\begin{alignat}{2}
B_{ij}&^{\beta\gamma \alpha}(t_1-t_2,t_1-t_3) =  \left\langle \left[u_j^{\beta}(t_1)-u_i^{\beta}(t_1) \right]\right. \notag \\
  &\left. \times \left[ u_j^{\gamma}(t_2)-u_i^{\gamma}(t_2) \right] \left[\vphantom{u_j^{\beta}(t_1)}  v_i^{\alpha}(t_3)+v_j^{\alpha}(t_3) \right] \right\rangle. \label{eq:Bij_def} 
\end{alignat}
Note that in contrast to using just a single correlation time as in previous equations, we have defined the correlation function \eqref{eq:Bij_def} in a more general form as a function of two correlation times. While this is not necessary to calculate the spectral heat current, this definition proves useful in analyzing the elastic and inelastic scattering in more detail.
The correlation function $B_{ij}^{\beta\gamma\alpha}(0,\tau)$ appearing in Eq. \eqref{eq:Q_ABtau} can be written as $B_{ij}(0,\tau)= \int_{-\infty}^{\infty}(d\omega'/2\pi) \hat{B}_{ij}(\omega',\tau) $, where $\hat{B}$ is the Fourier transformation of $B$ with respect to the first time variable. Carrying out the Fourier transforms of $A$ and $\hat{B}$ with respect to $\tau$ in Eq. \eqref{eq:Q_ABtau} then allows for writing the Fourier transformed correlation function $\tilde K_{ij} (\omega)$ as
\begin{alignat}{2}
\tilde K_{ij} (\omega) \approx&  \frac{1}{2} \sum_{\alpha,\beta \in\{x,y,z\}}  k_{ij} ^{\alpha\beta} \tilde A^{\beta\alpha}_{ij}(\omega)\notag \\
&  +\frac{1}{4} \sum_{\alpha,\beta,\gamma \in\{x,y,z\}} \gamma_{ij}^{\alpha\beta\gamma} \int_{-\infty}^{\infty} \frac{d\omega'}{2\pi} \tilde B_{ij}^{\beta\gamma\alpha}(\omega',\omega). \label{eq:Kij_final}
\end{alignat}
Using Eqs. \eqref{eq:Gdef} and \eqref{eq:spectral_J}, summing over the particles $i \in I$ and $j\in J$ interacting across the interface and dividing by the interfacial temperature drop $\Delta T$ and area $A$, we finally get the spectral decomposition of the conductance between particle sets $I$ and $J$:
\begin{alignat}{2}
G &= \frac{2}{A\Delta T} \int_0^{\infty} \frac{d\omega}{2\pi} \sum_{\substack{i\in I \\ j\in J}} \textrm{Re}[\tilde{K}_{ij}(\omega)] \\
&\approx \int_0^{\infty} \frac{d\omega}{2\pi} \left[ g^{\textrm{el}}(\omega)+ g^{\textrm{inel}}(\omega) \right].\label{eq:Gomega}
\end{alignat}
Here the elastic part describing linear energy transfer processes at the interface is 
\begin{equation}
 g^{\textrm{el}}(\omega) = \frac{1}{A\Delta T}\sum_{\substack{i\in I \\ j\in J}} \sum_{\alpha,\beta \in\{x,y,z\}}  k_{ij} ^{\alpha\beta} \textrm{Re}\left[\tilde A^{\beta\alpha}_{ij}(\omega)\right]\label{eq:gel}
\end{equation}
and the part describing the contribution of first-order inelastic processes is the frequency integral
 \begin{equation}
 g^{\textrm{inel}}(\omega) = \int_{-\infty}^{\infty} \frac{d\omega'}{2\pi} \mathfrak{g}^{\textrm{inel}}(\omega,\omega'), \label{eq:ginel_dec}
\end{equation}
where 
\begin{equation}
 \mathfrak{g}^{\textrm{inel}}(\omega,\omega') = \frac{1}{2A\Delta T}\sum_{\substack{i\in I \\ j\in J}} \sum_{\alpha,\beta,\gamma \in\{x,y,z\}}  \gamma_{ij} ^{\alpha\beta\gamma} \textrm{Re}\left[\tilde B^{\beta\gamma\alpha}_{ij}(\omega',\omega)\right]. \label{eq:ginel}
\end{equation}
Equation \eqref{eq:Gomega} with the definitions \eqref{eq:gel}, \eqref{eq:ginel_dec}, and \eqref{eq:ginel} is our final result for the spectral decomposition of the thermal conductance $G$ and its elastic and inelastic contributions. The only approximation made in deriving Eq. \eqref{eq:Gomega} is the truncation of the interfacial force expansion \eqref{eq:f_expansion} after second order. The accuracy of the expansion can therefore be straightforwardly refined by including higher-order terms, if necessary.

\begin{figure}[tb!]
 \begin{center}
  \includegraphics[width=8.6cm]{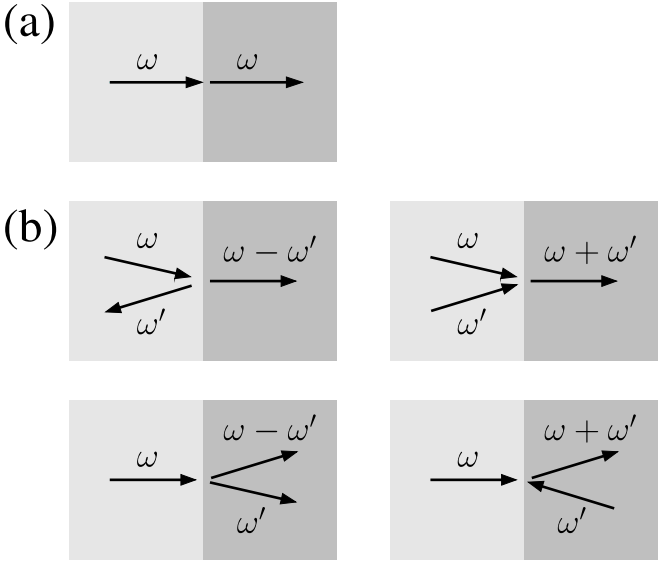}
  \caption{(a) Elastic and (b) inelastic thermal conduction processes at the interface, described by Eqs. \eqref{eq:gel} and \eqref{eq:ginel}, respectively. The four inelastic processes shown in (b) constitute only a subset of all processes obtained by expanding the parentheses on the left-hand side of Eq. \eqref{eq:B_fourier}.}  \label{fig:mechanisms}
 \end{center}
\end{figure}

In the general time-dependent case, the correlation function \eqref{eq:Jij1} would depend independently on both time variables $t_1$ and $t_2$ instead of just the difference $t_1-t_2$. In this case, the spectral heat current $q(\omega)$ would be a function of two frequency variables $\omega$ and $\omega'$, describing both energy-conserving ($\omega=\omega'$) and inelastic ($\omega\neq \omega'$) energy transfer. The assumption of steady state and thus translational invariance in time leads, however, to energy conservation, which is apparent when we write the general correlation functions in the right-hand sides of Eqs. \eqref{eq:Adef} and \eqref{eq:Bij_def} as
\begin{alignat}{2}
\frac{i}{\omega'}  &\left\langle \left[ \tilde v_j^{\beta}(\omega')- \tilde v_i^{\beta}(\omega') \right] \left[\vphantom{\tilde v_j^{\beta}(\omega')} \tilde v_i^{\alpha}(\omega)^*+ \tilde v_j^{\alpha}(\omega) ^*\right] \right\rangle \notag \\
 &= 2\pi \delta(\omega'-\omega) \tilde A^{\beta\alpha}_{ij}(\omega) \label{eq:A_fourier} 
\end{alignat}
and
\begin{alignat}{2}
& \frac{1}{\omega''\omega'} \left\langle \left[ \tilde v_j^{\beta}(\omega'')- \tilde v_i^{\beta}(\omega'') \right] \left[ \tilde  v_j^{\gamma}(\omega')^*- \tilde  v_i^{\gamma}(\omega')^* \right]  \right. \notag \\
 & \left. \times \left[\vphantom{\tilde v_j^{\beta}(\omega')} \tilde  v_i^{\alpha}(\omega)^*+ \tilde v_j^{\alpha}(\omega) ^*\right] \right\rangle = 2\pi \delta(\omega''-\omega-\omega') \tilde B^{\beta\gamma\alpha}_{ij}(\omega',\omega). \label{eq:B_fourier} 
\end{alignat}
Here we have used the identity
$
 \langle v_i(\omega) v_j(\omega')^* \rangle = 2\pi \delta(\omega-\omega') \int_{-\infty}^{\infty} dte^{i\omega t} \langle v_i(t) v_j(0)\rangle,
$
valid in the steady state when $\langle v_i(t+t_0) v_j(t_0)\rangle = \langle v_i(t)v_j(0)\rangle$ for any $t_0\in \mathbb{R}$, and also written the displacements $\tilde{u}_i^{\alpha}(\omega)$ and $\tilde{u}_j^{\alpha}(\omega)$ in terms of velocities $\tilde v_i^{\alpha}(\omega)$ and $\tilde v_j^{\alpha}(\omega)$ using the general relation $\tilde v_j^{\beta}(\omega')=-i\omega' \tilde u_j^{\beta}(\omega')$. The Dirac delta functions on the right-hand sides of Eqs. \eqref{eq:A_fourier} and \eqref{eq:B_fourier} ensure overall energy conservation. By expanding the parentheses on the left-hand side of Eq. \eqref{eq:A_fourier}, one can identify terms describing the coupling of vibrations at the two sides of the interface. For example, the term $ \langle \tilde v_i^{\beta} (\omega')\tilde v_j^{\alpha}(\omega)^*\rangle$ appearing on the left-hand side of Eq. \eqref{eq:A_fourier} can be interpreted as direct energy transfer between sites $i$ and $j$, mediated by the force constant $k_{ij}^{\alpha\beta}$ multiplying $\tilde A_{ij}^{\beta\alpha}$ in Eq. \eqref{eq:Kij_final}. This process is schematically depicted in Fig. \ref{fig:mechanisms}(a).

Similarly, the left-hand side of Eq. \eqref{eq:B_fourier} describes three-site processes in which atoms vibrating at frequencies $\omega$ and $\omega'$ create a vibration at frequency $\omega''=\omega+\omega'$, enabled by the first-order anharmonic force constant $\gamma_{ij}^{\alpha\beta\gamma}$. All the combinations of such three-vibration interactions at different sides of the interface are included in the eight terms obtained by expanding the parentheses in Eq. \eqref{eq:B_fourier}. Some of these inelastic emission and absorption processes are schematically depicted in Fig. \ref{fig:mechanisms}(b). 

\section{Numerical results: Elastic and inelastic energy transmission between mass-mismatched Lennard-Jones solids}
\label{sec:results}

\subsection{Structure}

\begin{figure}[t]
 \begin{center}
  \includegraphics[width=8.6cm]{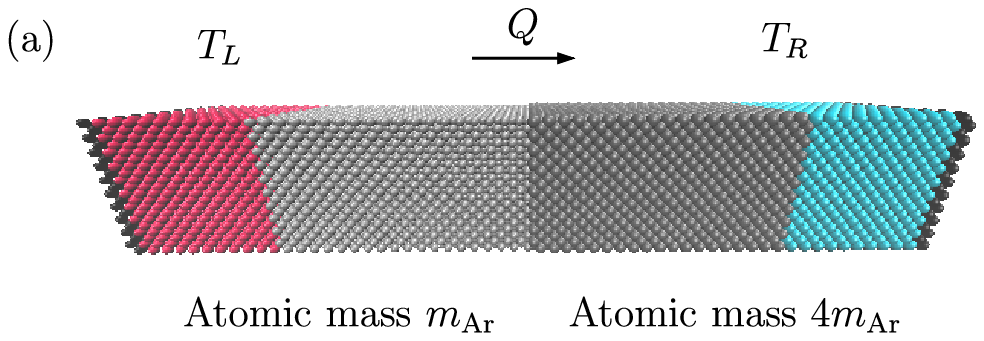}
  \includegraphics[width=8.6cm]{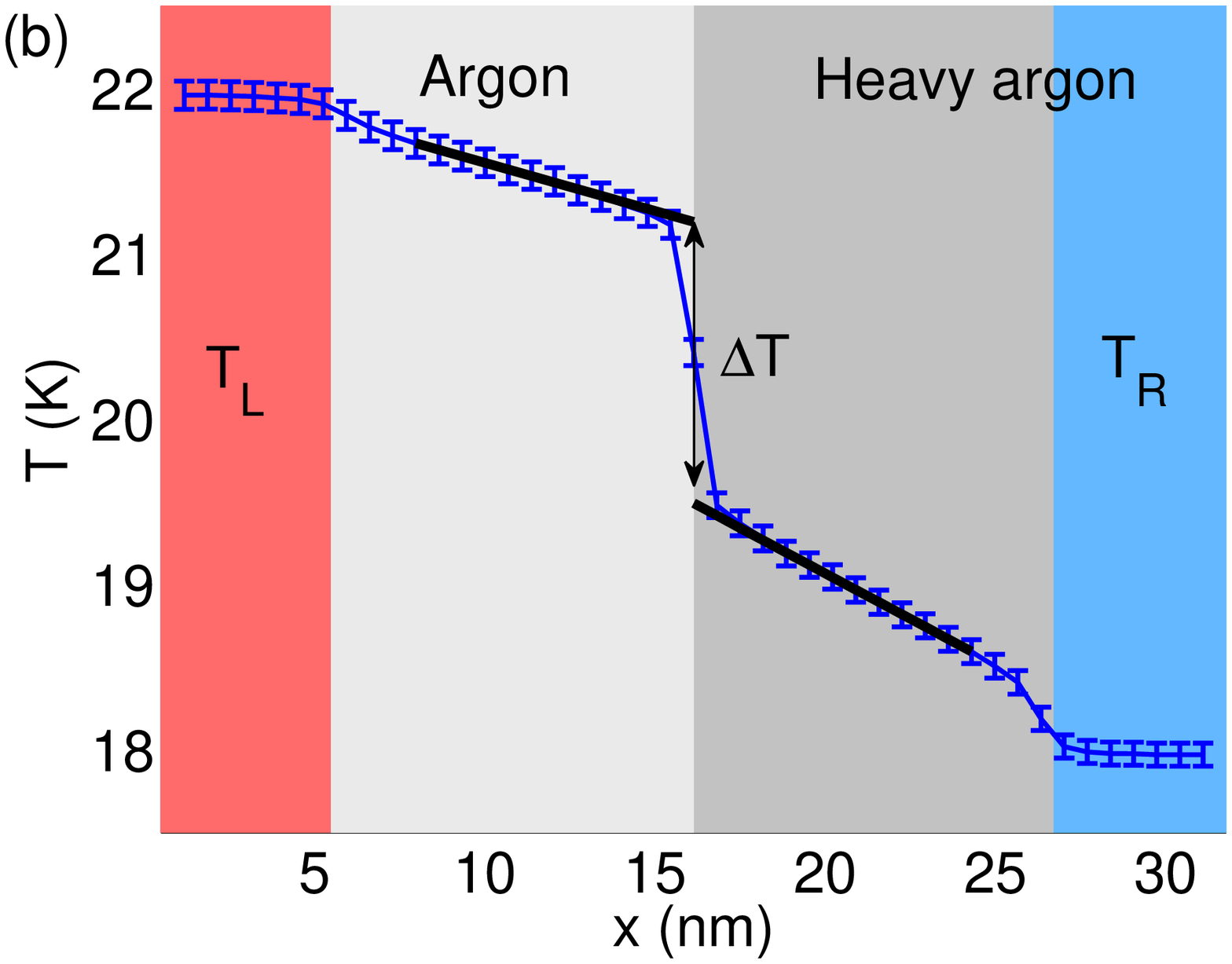}
  \caption{(Color online) (a) Atomistic illustration of the studied interface between two mass-mismatched Lennard-Jones solids. The atoms at the left and right ends are coupled to Langevin heat baths at different temperatures $T_L$ and $T_R$ to drive thermal current $Q$ through the interface in the middle. (b) Local kinetic temperature profile in a non-equilibrium simulation with average temperature $T=20$ K. Temperature drop $\Delta T$ at the interface is estimated by extrapolating the linear fits to the temperature profiles at different sides of the interface and calculating the difference at the interface. }  \label{fig:systems}
 \end{center}
\end{figure}
To obtain insight into the linear and non-linear energy transfer mechanisms across boundaries between two dissimilar solids, we calculate the spectral interfacial conductances \eqref{eq:gel} and \eqref{eq:ginel} between two Lennard-Jones (LJ) face-centered cubic (fcc) lattices illustrated in Fig. \ref{fig:systems}(a). The primary motivation for applying the method to LJ solids instead of, say, more realistic Si/Ge interface studied in numerous earlier works \cite{landry09,tian12,chalopin13}, is the strong non-linearity of the LJ potential, which makes the harmonic Green's function method insufficient and also allows for the suppression of finite-size effects already in relatively small simulation domains. The LJ potential also simplifies the theoretical discussion, as we do not have to consider the contribution of three-body forces or optical phonons on thermal conduction. If necessary, these can be straightforwardly investigated using the same formalism. The potential can also capture two key features of general pair potentials: the strong repulsion at short distances and weaker attraction at large distances. We therefore expect that our conclusions regarding the inelastic effects remain valid also for more complicated pair-potentials. The face-centered cubic lattice has been used as the lattice model in numerous earlier works investigating the effects of various material parameters on interfacial transport, including lattice constant\cite{pettersson90} and mass\cite{merabia12} mismatch, the strength of interfacial bonding \cite{shen11}, low-dimensional contacts \cite{panzer08}, interface roughness\cite{fagas99,daly02,stevens07,zhao09} and phonon-mediating thin films\cite{english12}. 


The velocity trajectories required for the determination of the Fourier transformed correlation functions \eqref{eq:A_fourier} and \eqref{eq:B_fourier} appearing in Eqs. \eqref{eq:gel} and \eqref{eq:ginel} are extracted from non-equilibrium steady-state simulations using the LAMMPS molecular dynamics simulation software \cite{plimpton95,lammps_website}. The parameters of the LJ pair-potential $V(r)=4\varepsilon[(\sigma_{\textrm{LJ}}/r)^{12}-(\sigma_{\textrm{LJ}}/r)^6]$, which we choose to correspond to solid argon, are \cite{allentildesley} $\varepsilon=1.67\times 10^{-21}$ J and $\sigma_{\textrm{LJ}}=3.4$ \AA. The potential cut-off distance is $r_c=2.5\sigma_{\textrm{LJ}}$ and the time-step in velocity Verlet integration is $\Delta t=$ 4.3 fs as in Ref. \cite{merabia12}. Velocity trajectories are collected from a MD run of $2\times 10^7$ steps, corresponding to 85.6 ns of physical time. Since no interactions take place beyond the cut-off distance $r_c$, only atoms located within the cut-off distance from the interface need to be included in the particle sets $I$ and $J$ consisting of the atoms located to the left and right of the interface, respectively. To introduce acoustic mismatch at the fcc (100) interface between the two LJ media, we set the masses of the atoms on the left and right sides of the interface to be $m_1=39.948$ amu and $m_2=4m_1$, where $m_1$ is the mass of argon. The mass-mismatch introduces a mismatch in the densities of vibrational states in the two materials, resulting in non-zero interfacial resistance. 

The total length of our simulated system is 60 cubic unit cells corresponding to the physical length $L=60a$, where $a$ is the fcc lattice constant changing between $1.5496\sigma_{\textrm{LJ}}$ at $T=0$ K and $1.58\sigma_{\textrm{LJ}}$ at $T=40$ K. Periodic boundary conditions are imposed in the direction parallel to the interface plane (transverse to the current flow). The width of the simulation area in these transverse directions is ten unit cells, corresponding to the cross-section area $A=100a^2$. Two monolayers of atoms at both ends of the system are fixed to avoid atomic sublimation. Twenty monolayers of atoms (corresponding to the physical length $L_{\textrm{bath}}=10a$) next to the fixed atoms at the left and right ends of the structure are coupled to Langevin heat baths at temperatures $T_L=T+\Delta T_b/2$ and $T_R=T-\Delta T_b/2$, and the temperature bias is chosen to be $\Delta T_b=T/5$. The bath time constant is chosen as $t_{\textrm{bath}}=2.14$ ps, ensuring that the bath-induced mean free path of phonons \cite{li09jap,saaskilahti13} satisfies $\Lambda_{\textrm{bath}}=c_s t_{\textrm{bath}}\lesssim L_{\textrm{bath}}$ ($c_s\approx 1250$ m/s is the speed of sound) so that phonons arriving at thermalized regions are dissipated before reflecting from the fixed ends. This choice ensures that the shown results are not sensitive to an increase in the length of the system, which we have also carefully checked.   

A typical local temperature profile obtained from the non-equilibrium simulation is shown in Fig. \ref{fig:systems}(b). The temperature drop $\Delta T$ at the interface is estimated from the difference of the linear temperature profiles extrapolated to the interface as illustrated in Fig. \ref{fig:systems}(b). This definition delivers an unambiguous $\Delta T$ and does not require choosing which atoms to include in the microscopic temperature calculation. The precise definition is not, however, important for our purposes, as $\Delta T$ only operates as a constant scaling factor in the spectral conductance distributions. We also note that the relative temperature drop $\Delta T/\Delta T_b$ at the interface (not shown) decreases from $0.91$ at $T=0$ K to approximately $0.3$ at $T=30$ K as a function of temperature, because higher temperature (i) reduces the interfacial resistance (see below) and (ii) increases the thermal gradient in the bulk by decreasing thermal conductivity. The temperature drop at low temperature agrees with the value calculated from ballistic Landauer-B\"uttiker formalism \cite{nazarov}: $\Delta T=R\Delta T_b$, where $R\approx 0.91$ is the interface reflectivity.

To ensure that the heat flow is in the linear regime, we have checked that the shown spectral conductances remain unchanged when the bias $\Delta T_b$ is halved. Smaller bias requires, however, longer simulation runs for retaining the same level of statistical accuracy. A very small non-linear effect can be observed in Fig. \ref{fig:systems}(b), where the average temperature at the interface slightly differs from the average bath temperature 20 K due to the combined effect of the asymmetry in the masses and temperature-induced nonlinear dynamics. 
 



\subsection{Elastic and inelastic energy transmission}

\begin{figure}[htb!]
 \begin{center}
 \includegraphics[width=8.6cm]{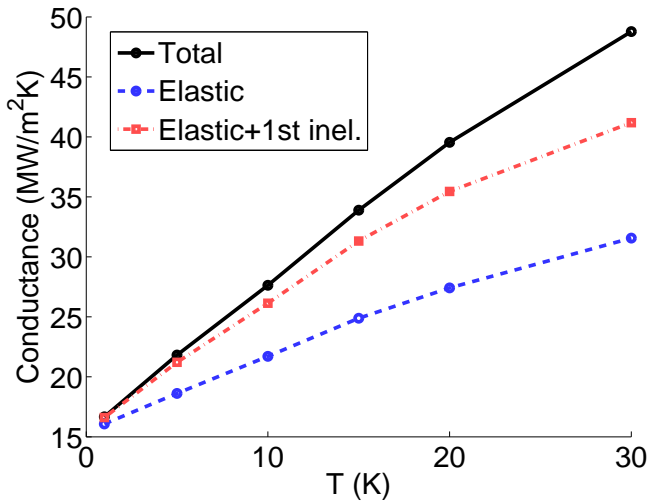}
   \caption{(Color online) Interfacial conductance as a function of temperature for the mass-mismatched Lennard-Jones interface. The total conductance is $G=Q/(A\Delta T)$, where $Q$ is the energy input (output) in the left (right) heat bath. The elastic conductance is $\int_0^{\infty} g^{\textrm{el}}(\omega)d\omega/(2\pi) $ [Eq. \eqref{eq:gel}] and the first-order inelastic addition to the elastic conductance is $\int_0^{\infty} g^{\textrm{inel}}(\omega)d\omega/(2\pi)  $ [Eq. \eqref{eq:ginel_dec}]. Only elastic transmission processes contribute to the conductance at very low temperature, but the importance of inelastic processes quickly increases with increasing temperature. Temperature change $\Delta T$ at the interface has been calculated as described in the caption of Fig. \ref{fig:systems}.} 
 \label{fig:GT_fcc}
 \end{center}
\end{figure}

To quantify the importance of inelastic processes to the interfacial conductance, we first plot the total interfacial conductance and its elastic and inelastic contributions as a function of temperature in Fig. \ref{fig:GT_fcc}. The total conductance $G=Q/(A\Delta T)$ (solid line), where the total heat current $Q$ includes all energy transmission mechanisms across the interface, has been determined from the work done by the heat baths on the atoms at the hot end of the structure. Below $T=20$ K, the conductance $G$ increases linearly with temperature as reported earlier \cite{stevens07}. At $T$ very close to $0$ K, linear approximation to the force in Eq. \eqref{eq:f_expansion} is accurate, and the elastic conductance obtained by integrating Eq. \eqref{eq:gel} over frequency agrees with the total conductance very well. As the temperature is increased, linear approximation to the force becomes insufficient very quickly, resulting in the under-estimation of the conductance. The inclusion of the first-order inelastic terms in the force remedies the under-estimation of the conductance reasonably well up to $T\sim 10 $ K, above which even higher-order processes start to contribute to the interfacial conductance. The spectral resolution of these higher-order processes corresponding to, e.g., four-phonon interactions at the interface is computationally challenging, so we limit our studies of interfacial inelastic processes to the first order (three-phonon processes). Because higher-order phonon processes are mainly responsible for the increasing thermal conductance at high temperatures, we have restricted the studied temperature range below $T=30$ K. In addition, increasing the temperature produces large thermal fluctuations in the interfacial heat current, leading to large statistical uncertainty in the conductance. Therefore, longer simulation runs would be required at high temperatures. 

\begin{figure}[htb!]
 \begin{center}
  \includegraphics[width=8.0cm]{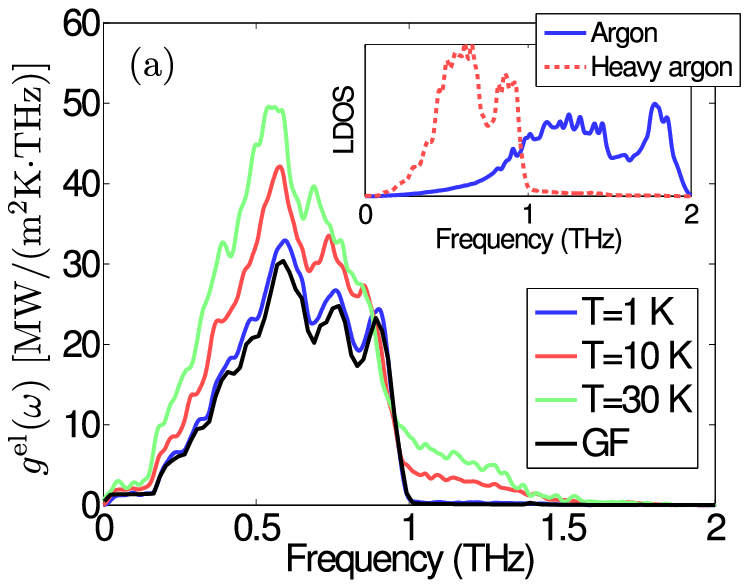}
  \includegraphics[width=8.0cm]{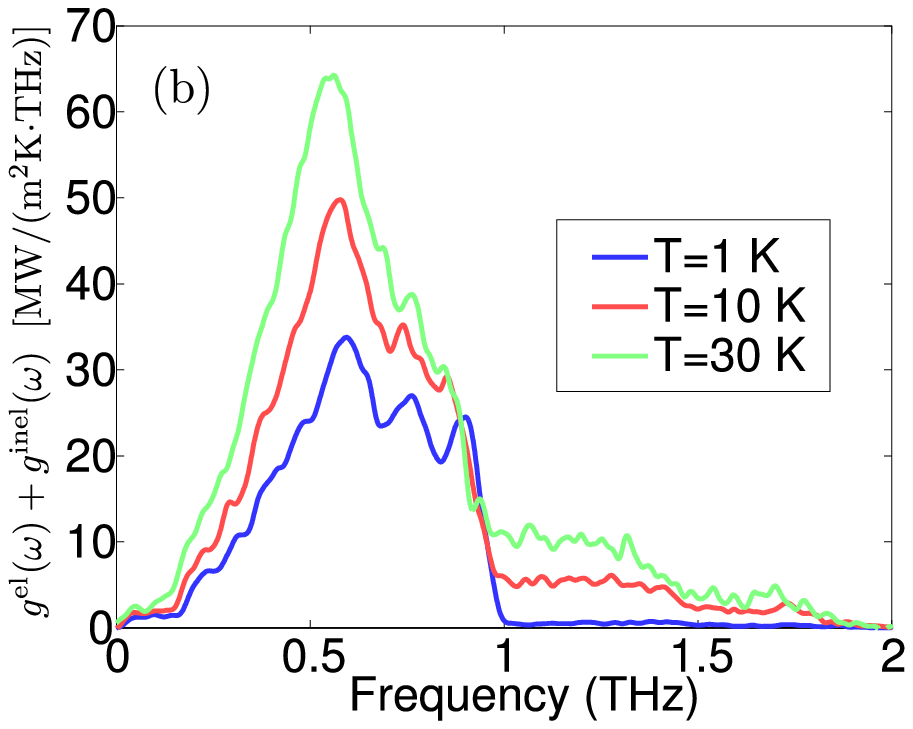}
   \caption{(a) The elastic conductance  \eqref{eq:gel} as a function of frequency at various temperatures. At $T=1$ K, the elastic conductance agrees very well with the Landauer-B\"uttiker conductance $g^{\textrm{LB}} (\omega)=k_B \mathcal{T}(\omega)/A$, where the transmission function $\mathcal{T}(\omega)$ has been calculated for the interface between two semi-infinite LJ solids using the Green's function (GF) method. At high temperatures, inelastic effects in the bulk enable energy transmission also above the cut-off frequency $f_c^{(2)}=1$ THz of the heavier solid. Inset: Local density of vibrational states (LDOS, arbitrary units) at the interface calculated from MD at $T=1$ K. In the lighter solid with mass $m_1=m_{\textrm{Ar}}$ (argon, solid line), vibration frequency cut-off is $f_c^{(1)}=2$ THz. In the heavier medium with mass $m_2=4m_1$ (heavy argon, dashed line), bulk vibrations therefore only extend up to $f_c^{(2)}=1$ THz, limiting ballistic transmission of phonons below this limit. At the interface, however, there are evanescent wave states extending up to 1.5 THz. (b) The sum $g^{\textrm{el}}(\omega)+ g^{\textrm{inel}}(\omega)$ of elastic and inelastic [Eq. \eqref{eq:ginel_dec}] spectral conductance as a function of frequency. At high temperatures, the inelastic energy transfer processes strongly enhance interfacial heat transfer at $f\approx 0.5$ THz and above the cut-off $f_c^{(2)}=1$ THz.} 
 \label{fig:T_fcc}
 \end{center}
\end{figure}

Figure \ref{fig:T_fcc}(a) shows the elastic conductance $g^{\textrm{el}}(\omega)$ [Eq. \eqref{eq:gel}] at the LJ interface at various temperatures. At low temperature ($T=1$ K), the elastic transmission agrees very well with the classical Landauer-B\"uttiker conductance $g^{\textrm{LB}}(\omega)=k_B\mathcal{T}(\omega)/A$, where the transmission function $\mathcal{T}(\omega)$ has been calculated for an interface between two semi-infinite LJ solids using the Green's function (GF) method \cite{zhang07,saaskilahti13}. The conductance plateau at small frequencies arises from the finite width $W\approx 5.3$ nm of the simulated interface and extends up to $f=c_s/W\approx 0.2$ THz, where the first transverse mode with non-zero transverse wave vector $k_y=2\pi/W$ can be excited. The value of the GF transmission per polarization at the plateau is equal to $\mathcal{T}(\omega)/3=8/9$, which agrees with the acoustic mismatch transmission factor \cite{swartz89} $\mathcal{T}_{\textrm{AMM}}=4Z_1Z_2/(Z_1+Z_2)^2=8/9$ for impedance mismatch $Z_2/Z_1=2$. For larger cross-sections, the transmission function smoothens and the low-temperature conductance, proportional to the number of excitable transverse modes, increases as $\omega^2$. We have carefully verified that increasing the system width $W$ to be larger than the presently used $W=10a$ only smoothens the steps in the spectral conductance profiles and does not affect any of the conclusions. 

In the low-temperature limit, only phonons below the cut-off frequency $f_c^{(2)}=1$ THz of the heavier material can carry heat across the interface, because higher frequency modes cannot propagate in the heavy argon. The reflected modes at the interface induce, however, evanescent vibrations, which can be observed in the interfacial density of states [inset of Fig. \ref{fig:T_fcc}(a)]. Due to the absence of any dissipation mechanism at low temperature, these vibrations cannot propagate into the bulk. As the temperature is increased, however, non-linearities in the soft LJ potential enable inelastic phonon-phonon interactions in the bulk in the vicinity of the interface, and consequently also phonons with frequencies above 1 THz can transmit their energy across the interface, as seen in Fig. \ref{fig:T_fcc}(a) for $T=10$ K and $T=30 $ K. This energy transfer above the frequency cut-off is dependent on inelastic phonon-phonon scattering \textit{in the bulk}, which is present in all the simulations and in the velocity statistics, but the actual energy transmission mechanism across the interface itself is still linear in Fig. \ref{fig:T_fcc}(a).  

Figure \ref{fig:T_fcc}(b) shows the sum $g^{\textrm{el}}(\omega)+ g^{\textrm{inel}}(\omega)$ of elastic and inelastic [Eq. \eqref{eq:ginel_dec}] spectral conductances as a function of frequency. In constrast to Fig. \ref{fig:T_fcc}(a), this also accounts for the contribution of \textit{anharmonic energy transfer} processes carrying heat across the interface. Compared to Fig. \ref{fig:T_fcc}(a), the non-linear interfacial interactions can be seen to strongly enhance the energy transfer at high temperatures, especially at $f\approx 0.5$ THz and above the frequency cutoff $f_c^{(2)}=1$ THz of the heavier material. Whereas linear interfacial interactions enable energy transfer only up to $f\approx 1.5$ THz at $T=30$ K, the inelastic interfacial processes can be seen to facilitate energy transfer up to the cut-off $f_c^{(1)} =2$ THz. 

\begin{figure}[tb]
 \begin{center}
  \includegraphics[width=8.6cm]{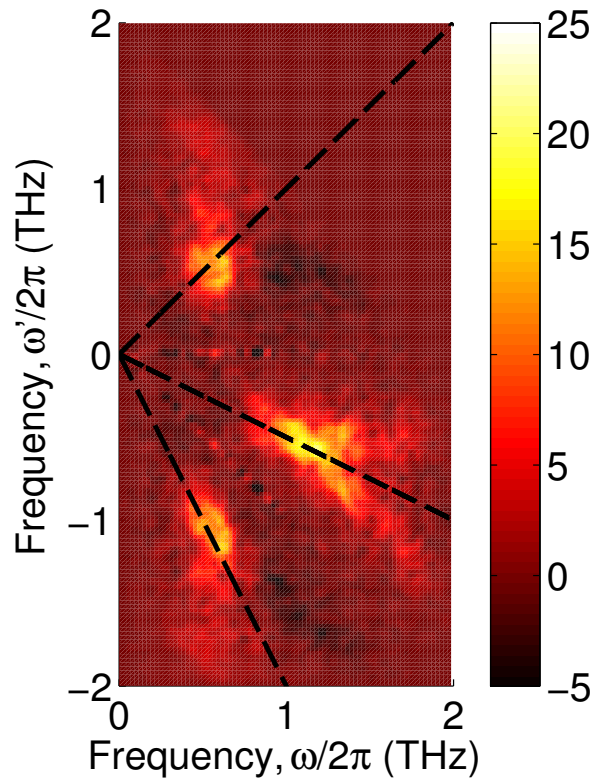}
   \caption{(Color online) Map of the inelastic three-phonon contribution $\mathfrak{g}^{\textrm{inel}}(\omega,\omega')$ [MW/(m$^2$K$\cdot$THz$^2$)] to the interfacial conductance at $T=20$ K. Processes satisfying either $\omega'=\omega$, $\omega'=-\omega/2$ or $\omega'=-2\omega$ (dashed lines) dominate the inelastic energy transfer.} 
 \label{fig:anharm_fcc}
 \end{center}
\end{figure}

Figure \ref{fig:anharm_fcc} shows the detailed two-dimensional map $\mathfrak{g}^{\textrm{inel}}(\omega,\omega')$ [Eq. \eqref{eq:ginel}] of inelastic phonon conductance across the interface at $T=20$ K. Frequencies $\omega$ and $\omega'$ correspond to the frequencies of two phonons participating in the process, the frequency of the third being fixed at $\omega''=\omega+\omega'$ by energy conservation [Eq. \eqref{eq:B_fourier}]. Figure shows that especially processes falling on lines $\omega'=\omega$, $\omega'=-\omega/2$ and $\omega'=-2\omega$ (dashed lines) dominate the inelastic contribution to the conductance. In these processes, the frequencies of the third phonon participating in the process are, respectively, $\omega''=2\omega$, $\omega''=\omega/2$ and $\omega''=-\omega$, implying that each process corresponds either to ''frequency-doubling'' or ''frequency-halving'' process at the interface. Such processes have been argued also earlier to dominate the inelastic energy transfer across the interface \cite{hopkins09_jap,hopkins11}, and the data in Fig. \ref{fig:anharm_fcc} strongly support this hypothesis. At temperatures lower than 20 K, the frequency maps are similar as in Fig. \ref{fig:anharm_fcc}, but the absolute values are scaled down in magnitude due to the reduced probability of anharmonic interactions. We have checked that the inelastic three-phonon contribution is very similar to Fig. 5 also for the mass ratio $m_2/m_1=2$ (not shown), suggesting that the above-mentioned processes dominate also more generally.

\subsection{Energy transfer under uniaxial pressure}

\begin{figure}[tb]
 \begin{center}
  \includegraphics[width=8.6cm]{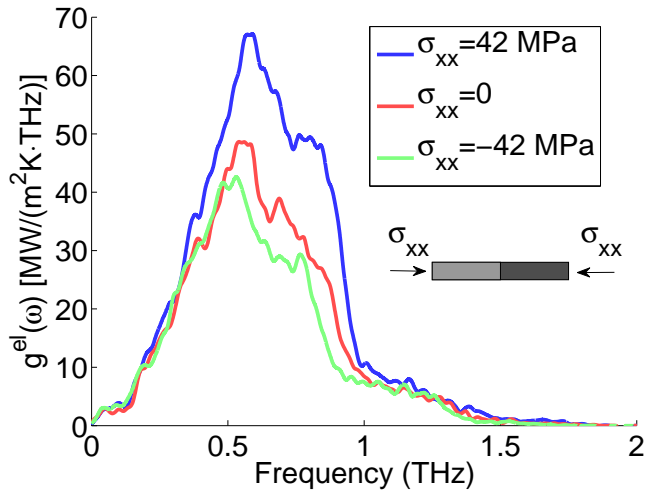}
   \caption{(Color online) Effect of uniaxial stress $\sigma_{xx}$ on the interfacial elastic conductance at $T=20$ K. Upon the application of compressive stress ($\sigma_{xx}>0$), the transmission of propagating phonons between 0.5 THz and the cut-off $f_c^{(2)}=1$ THz is enhanced. The conductance peak at $f\approx 0.6$ THz is also shifted to a slightly higher frequency due to lattice stiffening. Tensile stress ($\sigma_{xx}<0$), on the other hand, reduces phonon transmission at the interface due to lattice softening. Transmission of phonons at low frequencies ($f\lesssim0.5$ THz) and above the cut-off ($f\gtrsim 1$ THz) is not affected by the pressure.} 
 \label{fig:pressure}
 \end{center}
\end{figure}

Earlier MD studies have suggested that external pressure can be used to tune the interfacial conductance between dissimilar materials \cite{shen11}. At planar crystalline interfaces, the pressure modifies the interfacial bonding stiffness, which in turn affects the conductance. To get more insight into the effects of pressure on the conduction mechanisms at the interface, we show in Fig. \ref{fig:pressure} the elastic spectral conductance \eqref{eq:gel} for the LJ interface under compressive and tensile stress. In our simulation, the lattice is allowed to freely relax in the directions parallel to the interface so that the only non-zero component in the stress tensor $\sigma$ is $\sigma_{xx}$ (uniaxial stress).

Under compressive stress ($\sigma_{xx}=\varepsilon/\sigma_{\textrm{LJ}}^3=42$ MPa), the energy carried by phonons with frequencies between $0.5$ THz and $1$ THz can be seen to be strongly enhanced by the applied stress, implying that the transmission probability of phonons  through the interface is increased in this frequency range. Under tensile stress ($\sigma_{xx}<0$), the transmission is similarly reduced in the same frequency range due to lattice softening. This frequency range corresponds to phonons with mid-range wavelengths. 

At small frequencies $f\lesssim 0.5$ THz, the spectral conductance is nearly unaffected by the uniaxial stress. The small effect of stress on low frequency energy transmission can be understood by noting that the transmission probability of the long-wavelength phonons across the interface is close to unity even under zero stress and cannot be enhanced further by the application of the compressive stress as for mid-wavelength phonons, which have a smaller transmission probability. In addition, the long-wavelength transmission probability is determined by the ratio of the acoustic impedances, which is largely unaffected by the uniform stress.

The spectral conductance is independent of the stress also above the cut-off $f_c^{(2)}=1$ THz, where the non-zero transmission originally arises from the inelastic processes in the bulk (Fig. \ref{fig:T_fcc}). The weak dependence of the spectral conductance on the pressure above 1 THz implies that the inelastic processes are only weakly affected by the pressure. This can also be seen by evaluating the total conductance and analyzing the relative contributions of different mechanisms. For $\sigma_{xx}=42$ MPa, $\sigma_{xx}=0$ and $\sigma_{xx}=-42$ MPa, respectively, the total conductances $G=Q/(A\Delta T)$ are 48, 38 and 35 MW/(m$^2$K), showing that the conductance is strongly enhanced by the applied uniaxial stress. The contributions of the elastic conductance $G^{\textrm{el}}=\int_0^{\infty} g^{\textrm{el}}(\omega) d\omega/(2\pi)$ are, respectively, 37, 26 and 23 MW/(m$^2$K), so the contribution $G-G^{\textrm{el}}$ of all inelastic processes at the interface is nearly independent of stress, $G-G^{\textrm{el}}\approx 12$ MW/(m$^2$K). This shows that the elastic processes are essentially responsible for the observed enhancement of the interfacial conductance under compressive stress, while the inelastic processes are nearly unaffected. In disordered or weakly bonded interfaces, where pressure can cause deformations by the creation and breaking of bonds, the effect of pressure on interfacial conduction can, however, be much larger \cite{shen11,chalopin12}. 

The presently used MD model does not account for any quantum effects, which may affect our results at temperatures lower than the material's Debye temperature. In particular, classical dynamics overestimates the thermal occupation numbers of modes with energies higher than the thermal energy and may therefore slightly overestimate their contribution to interfacial conduction and the probability of inelastic scattering events. However, as our goal has been to investigate vibrational heat transfer mechanisms, detailed inclusion of quantum effects is not considered necessary, although quantum statistics could be partially included, e.g., by replacing the classical Langevin baths employed in this work by quantum thermal baths \cite{wang07,dammak09}.

For simplicity, we have only simulated perfectly smooth interfaces. At realistic interfaces, interfacial disorder is nearly always present and could have a strong impact on the interfacial heat transfer mechanisms. For example, we could expect that interfacial disorder broadens the strong conductance peaks in the inelastic conductance distribution (Fig. \ref{fig:anharm_fcc}) by introducing spatial incoherence in the interface scattering. Deeper investigation of such disorder effects is left for future work.



\section{Conclusions}
\label{sec:conclusions}

We investigated the contribution of anharmonic interactions and the resulting inelastic effects to interfacial thermal conduction at an interface between two mass-mismatched face-centered cubic lattices. The calculations were based on determining the spectral decomposition of interfacial heat current from dynamic correlation functions obtained from non-equilibrium MD simulations. At low temperatures, inelastic effects caused by the anharmonicity were negligible and the spectral conductance matched the results obtained from the harmonic Green's function method. As the temperature was increased, anharmonic effects became significant and facilitated energy transfer by the enhanced damping of evanescent vibrational modes close to the interface and three-phonon energy transfer processes at the interface. Spectral decomposition of the inelastic processes showed that frequency-doubling and frequency-halving processes dominated the three-phonon energy transfer. The results also revealed that for an interface under compressive uniaxial pressure, only the transmission of mid-wavelength phonons was enhanced by the pressure, while long-wavelength and inelastic energy transfer were nearly unaffected by the pressure.

The developed method for the spectral decomposition of thermal conductance provides substantial insight into the frequency dependence of inelastic scattering and Kapitza resistance between solid materials. Combined with the flexibility and versatility of molecular dynamics simulations, the presented method is applicable for detailed studies of a wide range of materials. We expect the method to prove highly useful in the computational optimization of interfaces, superlattices and even bulk materials for improving the efficiency of practical thermal devices.

\section{Acknowledgements}
We thank Shiyun Xiong, Haoxue Han, Yuriy Kosevich and Yann Chalopin for useful discussions. The computational resources were provided by the Finnish IT Center for Science, Aalto Science-IT project and Mesocentre de calcul de Centrale Paris. The work was partially funded by the Graduate School in Electronics, Telecommunication and Automation (GETA), the Condensed Matter and Materials Physics network (CMMP) of Aalto University and the Aalto Energy Efficiency Research Programme (AEF).


\begin{thebibliography}{10}

\bibitem{pop10}
E.~Pop,
\newblock Nano Res. {\bf 3}, 147 (2010).

\bibitem{cahill03}
D.~G. Cahill, W.~K. Ford, K.~E. Goodson, G.~D. Mahan, A.~Majumdar, H.~J. Maris,
  R.~Merlin, and S.~R. Phillpot,
\newblock J. Appl. Phys. {\bf 93}, 793 (2003).

\bibitem{cahill14}
D.~G. Cahill, P.~V. Braun, G.~Chen, D.~R. Clarke, S.~Fan, K.~E. Goodson,
  P.~Keblinski, W.~P. King, G.~D. Mahan, A.~Majumdar, H.~J. Maris, S.~R.
  Phillpot, E.~Pop, and L.~Shi,
\newblock Appl. Phys. Rev. {\bf 1},  (2014).

\bibitem{kapitza41}
P.~L. Kapitza,
\newblock J. Phys. USSR {\bf 4}, 181 (1941).

\bibitem{khalatnikov52}
I.~M. Khalatnikov,
\newblock Zh. Eksp. Teor. Fiz. {\bf 22}, 687 (1952).

\bibitem{little59}
W.~A. Little,
\newblock Canadian Journal of Physics {\bf 37}, 334 (1959).

\bibitem{swartz89}
E.~T. Swartz and R.~O. Pohl,
\newblock Rev. Mod. Phys. {\bf 61}, 605 (1989).

\bibitem{mingo03}
N.~Mingo and L.~Yang,
\newblock Phys. Rev. B {\bf 68}, 245406 (2003).

\bibitem{zhang07}
W.~Zhang, T.~S. Fisher, and N.~Mingo,
\newblock Numer. Heat Transfer, Part B {\bf 51}, 333 (2007).

\bibitem{shen11}
M.~Shen, W.~J. Evans, D.~Cahill, and P.~Keblinski,
\newblock Phys. Rev. B {\bf 84}, 195432 (2011).

\bibitem{chalopin12b}
Y.~Chalopin, K.~Esfarjani, A.~Henry, S.~Volz, and G.~Chen,
\newblock Phys. Rev. B {\bf 85}, 195302 (2012).

\bibitem{allentildesley}
M.~P. Allen and D.~J. Tildesley,
\newblock {\em Computer Simulation of Liquids} (Oxford University Press,
  Oxford, 2006).

\bibitem{puech86}
L.~Puech, G.~Bonfait, and B.~Castaing,
\newblock J. Low Temp. Phys. {\bf 62}, 315 (1986).

\bibitem{maiti97}
A.~Maiti, G.~Mahan, and S.~Pantelides,
\newblock Solid State Communications {\bf 102}, 517  (1997).

\bibitem{mullerplathe97}
F.~Müller-Plathe,
\newblock J. Chem. Phys {\bf 106}, 6082 (1997).

\bibitem{jund99}
P.~Jund and R.~Jullien,
\newblock Phys. Rev. B {\bf 59}, 13707 (1999).

\bibitem{barrat03}
J.-L. Barrat and F.~Chiaruttini,
\newblock Mol. Phys. {\bf 101}, 1605 (2003).

\bibitem{rajabpour10}
A.~Rajabpour and S.~Volz,
\newblock J. Appl. Phys. {\bf 108},  (2010).

\bibitem{chalopin12}
Y.~Chalopin, N.~Mingo, J.~Diao, D.~Srivastava, and S.~Volz,
\newblock Appl. Phys. Lett. {\bf 101},  (2012).

\bibitem{chalopin13}
Y.~Chalopin and S.~Volz,
\newblock Appl. Phys. Lett. {\bf 103},  (2013).

\bibitem{hardy63}
R.~J. Hardy,
\newblock Phys. Rev. {\bf 132}, 168 (1963).

\bibitem{lepri03}
S.~Lepri, R.~Livi, and A.~Politi,
\newblock Phys. Rep. {\bf 377}, 1 (2003).

\bibitem{saaskilahti12}
K.~S\"a\"askilahti, J.~Oksanen, R.~P. Linna, and J.~Tulkki,
\newblock Phys. Rev. E {\bf 86}, 031107 (2012).

\bibitem{landry09}
E.~S. Landry and A.~J.~H. McGaughey,
\newblock Phys. Rev. B {\bf 80}, 165304 (2009).

\bibitem{tian12}
Z.~Tian, K.~Esfarjani, and G.~Chen,
\newblock Phys. Rev. B {\bf 86}, 235304 (2012).

\bibitem{pettersson90}
S.~Pettersson and G.~D. Mahan,
\newblock Phys. Rev. B {\bf 42}, 7386 (1990).

\bibitem{merabia12}
S.~Merabia and K.~Termentzidis,
\newblock Phys. Rev. B {\bf 86}, 094303 (2012).

\bibitem{panzer08}
M.~A. Panzer and K.~E. Goodson,
\newblock J. Appl. Phys. {\bf 103}, 094301 (2008).

\bibitem{fagas99}
G.~Fagas, A.~G. Kozorezov, C.~J. Lambert, J.~K. Wigmore, A.~Peacock,
  A.~Poelaert, and R.~den Hartog,
\newblock Phys. Rev. B {\bf 60}, 6459 (1999).

\bibitem{daly02}
B.~C. Daly, H.~J. Maris, K.~Imamura, and S.~Tamura,
\newblock Phys. Rev. B {\bf 66}, 024301 (2002).

\bibitem{stevens07}
R.~J. Stevens, L.~V. Zhigilei, and P.~M. Norris,
\newblock International Journal of Heat and Mass Transfer {\bf 50}, 3977
  (2007).

\bibitem{zhao09}
H.~Zhao and J.~B. Freund,
\newblock J. Appl. Phys. {\bf 105},  (2009).

\bibitem{english12}
T.~S. English, J.~C. Duda, J.~L. Smoyer, D.~A. Jordan, P.~M. Norris, and L.~V.
  Zhigilei,
\newblock Phys. Rev. B {\bf 85}, 035438 (2012).

\bibitem{plimpton95}
S.~Plimpton,
\newblock J. Comp. Phys. {\bf 117}, 1  (1995).

\bibitem{lammps_website}
http://lammps.sandia.gov.

\bibitem{li09jap}
N.~Li and B.~Li,
\newblock J. Phys. Soc. Jpn. {\bf 78}, 044001 (2009).

\bibitem{saaskilahti13}
K.~S\"a\"askilahti, J.~Oksanen, and J.~Tulkki,
\newblock Phys. Rev. E {\bf 88}, 012128 (2013).

\bibitem{nazarov}
Y.~Nazarov and Y.~Blanter,
\newblock {\em Quantum Transport: Introduction to Nanoscience} (Cambridge
  University Press, 2009).

\bibitem{hopkins09_jap}
P.~E. Hopkins,
\newblock J. Appl. Phys. {\bf 106}, 013528 (2009).

\bibitem{hopkins11}
P.~E. Hopkins, P.~M. Norris, and J.~C. Duda,
\newblock J. Heat Transfer {\bf 133}, 062401 (2011).

\bibitem{wang07}
J.-S. Wang,
\newblock Phys. Rev. Lett. {\bf 99}, 160601 (2007).

\bibitem{dammak09}
H.~Dammak, Y.~Chalopin, M.~Laroche, M.~Hayoun, and J.-J. Greffet,
\newblock Phys. Rev. Lett. {\bf 103}, 190601 (2009).

\end{thebibliography}

\end{document}